\documentclass[prd,eqsecnum,nofootinbib]{revtex4}
\usepackage{bm}
\usepackage{amssymb}
\usepackage{graphicx}
\usepackage{epstopdf}
\usepackage{mathrsfs}
\usepackage{amsmath}
\numberwithin{equation}{part}
\begin{document}
\title{Aretakis Hair for Extreme Kerr Black Holes with Axisymmetric Scalar Perturbations}
\author{Lior M.~Burko$^{1}$, Gaurav Khanna$^{2,3,4}$ and Subir Sabharwal $^3$}
\affiliation{$^1$ Theiss Research, La Jolla, California 92037, USA \\
$^2$ Department of Physics, University of Rhode Island, Kingston, Rhode Island 02881, USA \\
$^3$ Center for Computational Research, University of Rhode Island, Kingston, Rhode Island 02881, USA \\
$^4$ Physics Department and Center for Scientific Computing \& Data Science Research, University of Massachusetts Dartmouth, Dartmouth, Massachusetts 02747, USA}
\date{April 12, 2023}
\begin{abstract} 
We study the evolution of axially-symmetric scalar field perturbations on an extreme Kerr spacetime for initial data with multipole moments $\ell^{\prime}$ higher than the least radiative mode, and we measure modes $\ell$ -- and for the first time also horizon charges -- that are excited by mode coupling interactions. We then find the Ori-Sela prefactors, a certain quantity that can be evaluated at finite distances and the Aretakis constant along the event horizon of the extreme Kerr black hole for a sequence of initial data preparations that differ only by their distance from the event horizon. We find that for initial data in the near field there is a linear relationship of the Aretakis constant and the Ori-Sela prefactor. For initial data farther than these the linear relationship is not universal, and we propose that stronger numerical simulations would be needed to regain linearity. The linear relationship suggests that the Aretakis charge along the event horizon can be measured at a finite distance, thereby extending this type of violation of the no-hair theorems from the least radiative axisymmetric mode also to situations that involve mode coupling. 

\end{abstract}
\maketitle

\section{Introduction} 
Extreme Reissner-Nordstr\"{o}m (ERN) black hole (BH) spacetimes exhibit a conformal symmetry \cite{Couch_Torrence:1984} that relates the Newman-Penrose constants at future null infinity ($\mathscr{I}^+$) with the Aretakis constants at the future event horizon (EH, $\mathscr{H}^+$) \cite{Bizon:2013,Lucietti:2013,Godazgar:2017,Bhattacharjee:2018}. This relationship suggests that at least for ERN one could at least in principle violate the no-hair theorems \cite{Bekenstein:1972} with measurements of Newman-Penrose constants at $\mathscr{I}^+$. Later, it was shown that one can indeed measure the Aretakis constants for ERN along $\mathscr{H}^+$ with measurements made at $\mathscr{I}^+$ \cite{Angelopoulos:2018,Burko:2019} and at finite distances \cite{Burko:2021}. (Note, that in \cite{Angelopoulos:2018} no use of the conformal symmetry was made.) In fact, \cite{Burko:2019,Burko:2021} also extended this result for extreme Kerr (EK) BHs, specifically for axisymmetric scalar and gravitational perturbations. 

The proposed external measurement of BH hair with Aretakis charges for EK is perhaps surprising, because the conformal symmetry of ERN does not extend to EK \cite{Couch_Torrence:1984}. However, it was pointed out in \cite{Bizon:2013} that axially symmetric scalar fields propagating on EK spacetimes do have such a conformal symmetry, a result closely related to the symmetry of the radial equation for such perturbations \cite{Couch_Torrence:1984}. Therefore, one may expect that at least in the axially symmetric case, although possibly not in general, one could still measure at finite distances Aretakis charges on $\mathscr{H}^+$, and thereby violate the no-hair theorems in this sense. 

In \cite{Burko:2021} we considered the case of the lowest radiative mode of a scalar field propagating on a fixed EK spacetime, specifically the axisymmetric monopole mode. That is, we excited in \cite{Burko:2021} the monopole mode, and then measured the Ori-Sela prefactor $e[\psi ]$ \cite{Ori:2013,Sela:2016} and the Aretakis constant for a set of initial data preparations differing only by their distance from the EK EH. We showed in \cite{Burko:2021} that there was a linear relationship between the two, such that measurement of the Ori-Sela prefactor at a finite distance could allow us to infer the Aretakis constant. We interpreted this measurement of the Aretakis constant from measurements made at a finite distance as a violation of the no-hair theorem. 

The Kerr spacetime, and specifically EK, exhibit an intricate mode coupling mechanism \cite{Burko:2014}. We therefore pose the question of whether the behavior shown in \cite{Burko:2021} for the lowest radiative mode persists also for modes that are excited by mode-coupling excitations.  
We study here the Aretakis charges and their measurements at finite distances for an initial $\ell'$ multipole mode of an axially symmetric massless scalar field that excites an $\ell$ multipole mode, $\,_{\ell'}\psi_{\ell}$. The latter gives rise to an Aretakis charge of degree $k$, $_{\ell^{\prime}}H_{k,\ell}$, of an EK, and we study its relationship to the generalized Ori-Sela prefactor, $_{\ell^{\prime}}e_{k,\ell}[\psi ]$. To the best of the knowledge of the present authors, this is the first time that horizon charges are calculated for mode that are created by mode coupling. 
By showing a linear relationship of the two we propose following \cite{Burko:2021} that one could at least in principle measure BH hair beyond those discussed in \cite{Bekenstein:1972} also for multipole modes beyond the least radiative mode. The BH hair we propose are a consequence of linear perturbation theory, and result from a (linear approximation) of dynamical processes. It remains an open question whether similar hairs can be found in the fully nonlinear theory. 

\section{Numerical approach}

We solve the scalar wave equation for perturbations in EK black hole backgrounds, focusing on axisymmetric modes ($m=0$). We modify the equation to work in compactified hyperboloidal coordinates $(\tau, \rho, \theta, \phi)$ that allow for time evolution on hypersurfaces which bring $\mathscr{I}^+$ to a finite radial coordinate $\rho(\mathscr{I}^+)=S<\infty$. The relationship between these new coordinates $(\tau,\rho)$ and the spherical Boyer-Lindquist coordinates $(t,r)$ is 
\begin{eqnarray}
\Omega &=& 1-\frac{\rho}{S}\nonumber \\
r &=& \frac{\rho}{\Omega(\rho)} \\
v:= t+r_*-r &=& \tau+\frac{\rho}{\Omega(\rho)}-\rho-4M\log\Omega(\rho)\nonumber
\end{eqnarray}
where $S$ denotes the location of $\mathscr{I}^+$ in hyperboloidal coordinates, $r_*$ is the usual `tortoise' coordinate and $v$ is the modified advanced time. Note that the angular variables are the same in both coordinate systems.

Our numerical implementation scheme entails re-writing the second order partial differential equation (PDE) in terms of two coupled first-order differential equations. We solve this system using a high-order weighted essentially non-oscillatory (WENO) finite-difference scheme with explicit Shu-Osher time-stepping. Details may be found in our previous work~\cite{code}. We choose $S=19.0$ and the location of $\mathscr{H}^+$ such that $\rho(\mathscr{H}^+)=0.95$. The initial data are a truncated Gaussian centered at $\rho=(1.0, 1.1, 1.2, 1.3, 1.4, 1.5)$ with a width of $0.22$ and non-zero for $\rho\in[0.95,8]$. This choice ensures compactly supported initial data but with non-zero support on the $\mathscr{H}^+$ surface. 

Finally, to complete these long duration, high-accuracy and high-precision computations in a reasonable time-frame we make extensive use GPGPU-based parallel computing. For additional details on implementation of such intensive computations on a parallel GPU architecture, we refer the reader to our earlier work on the subject~\cite{code}.

\section{Fall off rates at $\mathscr{I}^+$, $\mathscr{H}^+$, and at $r={\rm const}$}
We found before the fall-off rates for scalar perturbations ($s=0$) along $r={\rm const}$,  along $\mathscr{I}^+$, and along $\mathscr{H}^+$ for the case of no initial data supported on $\mathscr{H}^+$, and we add in Table \ref{table1} the corresponding decay rates when the initial data are supported on $\mathscr{H}^+$. We have extensive numerical support for the asymptotic decay rates that appear in Table \ref{table1}. The results in Table \ref{table1} allow us to predict the triplets $\ell',\ell;k$, where $k$ is the order of the Aretakis charge (which is related to the order of the transverse derivative operator along $\mathscr{H}^+$), that would produce Aretakis constants. 

\begin{figure}
\includegraphics[width=7.5cm]{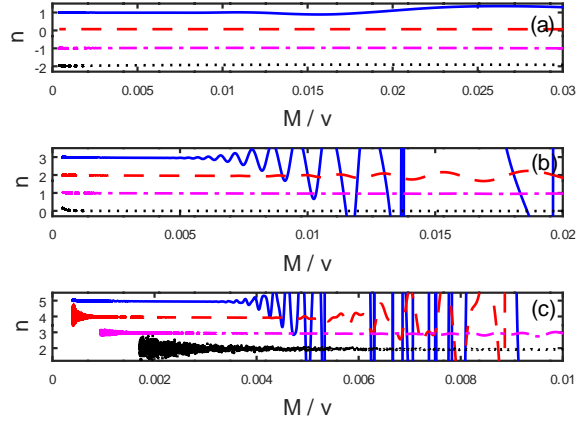}
\caption{The power law indices $n$ along $\mathscr{H}^+$ for the field $\psi\sim v^n$ (solid), and the transverse derivatives $\,\partial_u\psi$ (dashed), $\,\partial_u^2\psi$ (dash-dotted), and $\,\partial_u^2\psi$ (dotted), for $\ell'=2$ and $\ell=0$ (upper panel (a), $\ell=2$ (center panel (b)), and $\ell=4$ (lower panel (c)). The initial data have support on $\mathscr{H}^+$.}
\label{rates}
\end{figure}

\begin{figure}
\includegraphics[width=7.5cm]{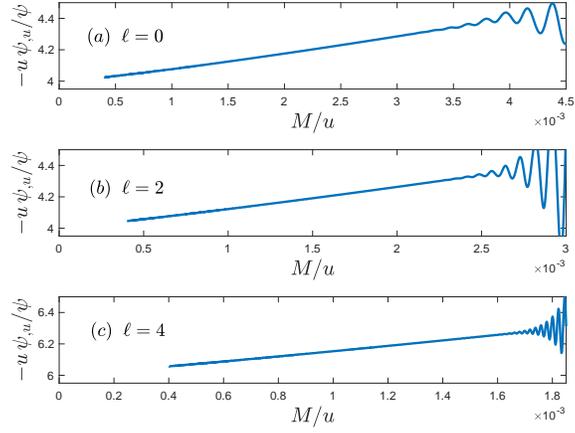}
\caption{The power law indices $n$ along $\mathscr{I}^+$ for the field $\psi\sim u^n$ (solid) for $\ell'=4$. Upper panel (a): $\ell=0$, center panel (b): $\ell=2$, and lower panel (c):  $\ell=4$. The initial data have support on $\mathscr{H}^+$.}
\label{scri_rates}
\end{figure}

\begin{table}[h]
\begin{tabular}{|c|c|c|}
\hline
 & Horizon Data & No Horizon Data \\
\hline\hline
$r={\rm const}$ & $-n=\left\{
\begin{matrix}
\ell^{\prime}+\ell+2\,\,\,\,\,\, , \ell'=0,1 \\
 \ell^{\prime}+\ell \,\,\,\,\,\, , {\rm otherwise}
\end{matrix}\right.$ & 
$-n=\left\{
\begin{matrix}
\ell^{\prime}+\ell+3\,\,\,\,\,\, , \ell'=0,1 \\
 \ell^{\prime}+\ell+ 1 \,\,\,\,\,\, , {\rm otherwise}
\end{matrix}\right. $
 \\
\hline
$\mathscr{I}^+$ & $-n=\left\{
\begin{matrix}
\ell^{\prime}\,\,\,\,\,\,\,\,\,\,\;\;\; , \ell\le\ell'-2 \\
\ell+2 \,\,\,\,\,\, , \ell\ge\ell^{\prime}
\end{matrix}\right.$ & $-n=\left\{
\begin{matrix}
\ell^{\prime}\,\,\,\,\,\,\,\,\,\,\;\;\; , \ell\le\ell'-2 \\
\ell+2 \,\,\,\,\,\, , \ell\ge\ell^{\prime}
\end{matrix}\right.$ \\
\hline
$\mathscr{H}^+$ & $-n=\left\{
\begin{matrix}
\ell^{\prime}-1\,\,\,\,\,\,\,\,\,\,\;\;\; , \ell\le\ell'-2 \\
\ell+1 \,\,\,\,\,\, , \ell\ge\ell^{\prime}
\end{matrix}\right.$ & $-n=\left\{
\begin{matrix}
\ell^{\prime}\,\,\,\,\,\,\,\,\,\,\;\;\; , \ell\le\ell'-2 \\
\ell+2 \,\,\,\,\,\, , \ell\ge\ell^{\prime}
\end{matrix}\right.$ \\
\hline
\end{tabular}
\caption{\label{table1} The power-law indices $n$ for late-time decay for $\psi$: Along $r={\rm const}$, $\psi(t)\sim t^{n}$; along $\mathscr{I}^+$,  $\psi(u)\sim u^{n}$; and along $\mathscr{H}^+$, $\psi(v)\sim v^{n}$. } 
\end{table}
The results in Table \ref{table1} were obtained from results such as Figs.~\ref{rates} and \ref{scri_rates}. 
Along $\mathscr{I}^+$ horizon data do not change the decay rate, and the latter is the same as without horizon data. This conclusion is consistent with Table 2 in \cite{Angelopoulos:2020}. The reason that with or without horizon data the decay rates along $\mathscr{I}^+$  are the same is that the initial data break the Couch-Torrence symmetry \cite{Couch_Torrence:1984}. 

We can use the results from Table \ref{table1} to find the power-law indices for transverse derivatives along $\mathscr{H}^+$. Specifically, without horizon data the $p^{\rm th}$ transverse derivative along $\mathscr{H}^+$, $\,\partial_u^p\psi(v)\sim v^{n}$ at late advanced times $v\gg M$, where 
\begin{eqnarray}
-n_{\rm no\; horizon\; data}=\left\{
\begin{matrix}
\ell^{\prime}-p\,\,\,\,\,\,\,\,\,\,\;\;\; , \ell\le\ell'-2 \\
\ell+2-p \,\,\,\,\,\, , \ell\ge\ell^{\prime}
\end{matrix}\right.
\label{A1}
\end{eqnarray}
and with horizon data 
\begin{eqnarray}
-n_{\rm horizon\; data}=\left\{
\begin{matrix}
\ell^{\prime}-1-p\,\,\,\,\,\,\,\,\,\,\;\;\; , \ell\le\ell'-2 \\
\ell+1-p \,\,\,\,\,\, , \ell\ge\ell^{\prime}
\end{matrix}\right. \, .
\label{A2}
\end{eqnarray}
We can use the results in Eqs.~(\ref{A1}) and (\ref{A2}) to predict at what value of $k$ we expect an Aretakis constant $_{\ell^{\prime}}H_{k,\ell}[\psi]$ given $\ell',\ell$. Specifically, setting $n=0$, we can solve Eq.~(\ref{A1}) for the derivative order $p$. Then the required $k$ is just $p-1$. Therefore, with no horizon data we expect
\begin{eqnarray}
k_{\rm no\; horizon\; data}=\left\{
\begin{matrix}
\ell^{\prime}-1\,\,\,\,\,\,\,\,\,\,\;\;\; , \ell\le\ell'-2 \\
\ell+1 \,\,\,\,\,\, , \ell\ge\ell^{\prime}
\end{matrix}\right.
\label{A3}
\end{eqnarray}
and with horizon data 
\begin{eqnarray}
k_{\rm horizon\; data}=\left\{
\begin{matrix}
\ell^{\prime}-2\,\,\,\,\,\,\,\,\,\,\;\;\; , \ell\le\ell'-2 \\
\ell \,\,\,\,\,\, , \ell\ge\ell^{\prime}
\end{matrix}\right. \, .
\label{A31}
\end{eqnarray}

Specific examples for the power law indices for different $\ell',\ell$ values and finding the $k$ corresponding to Aretakis charges are listed in Appendix A.

We find empirically that for $r>M$, outside the EH,  the radial profile for the dominant $\ell$-mode can be modeled by
\begin{equation}
_{\ell'}\psi_{\ell}\sim\, _{\ell'}e_{\ell}\,r^a\,(r-M)^b\,t^n\,\Theta_{\ell}(\theta)\, ,
\end{equation}
where
\begin{equation}
\begin{matrix}
a=1, b=-1  &  \ell^{\prime}\; {\rm is\; even} \\
a=1 , b=-2 &  \ell^{\prime}\; {\rm is\; odd} \\
\end{matrix} 
\end{equation}
and where $_{\ell'}e_{\ell}$ is the generalized Ori-Sela pre-factor. 

\section{Linear relationship of $e$ and $H$}

\begin{figure}
\includegraphics[width=7.5cm]{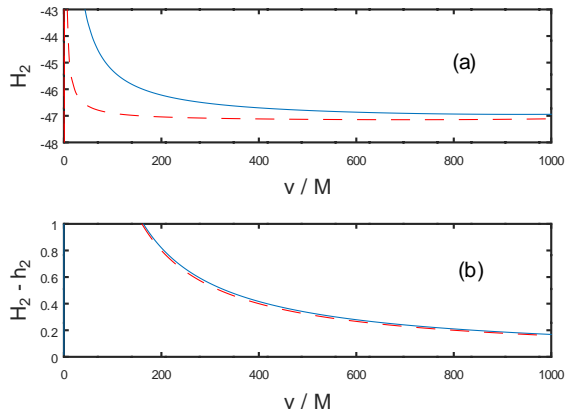}
\caption{The approximation of the Aretakis charge $_{\ell^{\prime}}H_{k,\ell}$ by $_{\ell^{\prime}}h_{k,\ell}$. Top panel (a): $_{2}H_{2,2}$ (solid) and $_{2}h_{2,2}$ (dashed). Bottom panel (b): $_{2}H_{2,2}-\,{_{2}h_{2,2}}$ (solid) and the reference curve $160M/v$ (dashed).}
\label{approx}
\end{figure}

We label the Aretakis constant $_{\ell^{\prime}}H_{k,\ell}$ where $k$ is related to the order of the differential operator, $\ell'$ is the multipole order of the perturbation field, and $\ell$ is the multipole order of the field of interest. Specifically, 
\begin{equation}
_{\ell^{\prime}}H_{k,\ell}[\psi]=\,\partial_r^{k+1}\left[r\,\partial_r\left(r\,_{\ell'}\psi_{\ell}\right)\right]
\end{equation} 
In practice, we approximate the Aretakis constant $_{\ell^{\prime}}H_{k,\ell}[\psi]$ with $_{\ell^{\prime}}h_{k,\ell}[\psi]$, where 
\begin{equation}
_{\ell^{\prime}}h_{k,\ell}[\psi]\sim M^2\,\partial_r^{k+1}\, _{\ell'}\psi_{\ell}
\end{equation} 
as is shown in Fig. \ref{approx}. 

Table \ref{T4} shows the values of $\ell'$ and $\ell$ for which we studied the relationship of the Aretakis charge $_{\ell^{\prime}}H_{k,\ell}$ and the Ori-Sela prefactor $_{\ell^{\prime}}e_{\ell}$. We find linear relationships $_{\ell^{\prime}}H_{k,\ell}=\beta\,_{\ell^{\prime}}e_{\ell'}+\alpha$ (see Figs.~\ref{000_022}, \ref{200_222}, and \ref{420_422}). See Appendix B for detail. In two of the cases studied we find deviations from linearity. Specifically, for $\ell'=0, \ell=2$ and for $\ell'=2,\ell=2$. These deviations from linearity occur when the initial data are far from the EH, but for near initial data the linear behavior is still observed. 
\begin{table}[h]
\begin{tabular}{|c||c|c|c|}
\hline
$\ell / \ell'$ & $\ell'=0$ & $\ell'=2$ & $\ell'=4$  \\
 \hline \hline
 $\ell=0$ & 0 & 0 & 2 \\
\hline
 $\ell=2$ & {\bf 2} & {\bf 2} & 2 \\
 \hline
\end{tabular}
\caption{The value of the order $k$ of the Aretakis charge $_{\ell^{\prime}}H_{k,\ell}$ for which a linear relationship to the Ori-Sela prefactor $_{\ell^{\prime}}e_{\ell}$ is found. In boldface we show the cases for which deviations from linearity are found. }
\label{T4}
\end{table}

In three of the cases studied (see Appendix B) we find that at the $95\%$ confidence level one cannot reject the claim that the intercept $\alpha=0$ with . We propose that more robust investigation may find this result to be a general rule.

Fully explaining these deviations from linearity is as yet an open question. We propose that more powerful numerical simulations would find linearity also for distant initial data: 
When plotting different $\ell$ projections as functions of $\rho$ for different sets of initial data (distinguished by the location of the peak) we find that up excitations behave differently for different initial data sets (and also for $\ell'=\ell$ when $\ell$ is not the lowest radiative mode), but the behavior is the same for $\ell=\ell'$ (when $\ell$ is the lowest radiative mode). This conclusion suggests that higher excitations may take longer to settle for far out initial data sets. This idea is strengthened by noticing that all deviations from linearity occur with $|_{\ell'}H_{k,\ell}|$ being under-valued, never over-valued. We cautiously propose that the dominant mode has saturated, but subdominant modes have not saturated yet, and therefore their contributions to $_{\ell'}H_{k,\ell}$ are not full. 

\begin{figure}
\includegraphics[width=7.5cm]{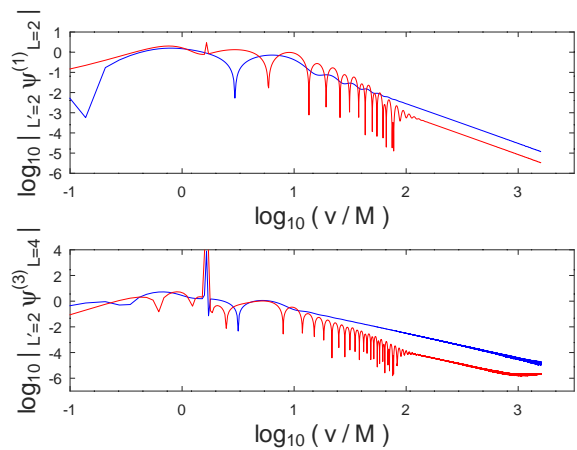}
\caption{Comparison of the behavior of subdominant modes. Top panel: $_2\psi^{(1)}_{2}$ for close initial data (upper curve at late times) and for far initial data (lower curve at late times). Bottom panel: $_2\psi^{(3)}_{4}$ for close initial data (upper curve at late times) and for far initial data (lower curve at late times). }
\label{compare}
\end{figure}

To test this idea we compare the contribution of subdominant modes to $_2H_{2,2}$ (nonlinear deviations) and to $_2H_{00}$ (no deviations from linearity). In the former case we take the subdominant mode $\ell'=2$, $\ell=4$ (up excitation), and in the latter case we take the subdominant mode $\ell'=2$, $\ell=2$ (up excitation). We find the results in Fig.~\ref{compare}. The deviations from power law behavior for $_2\psi^{(3)}_{4}$ at late time suggest that we do not get an accurate determination of $_2H_{2,2}$ which could explain the deviations shown in Fig.~\ref{200_222}(b). 

We comment that before the deviation from linearity in Fig.~\ref{200_222}(b) starts, asymptotic behavior is observed. Perhaps we need to evaluate the Aretakis constant in that domain, before presumably numerical effects change the behavior. If this is right, it is possible that one could still read the Aretakis constant from measurements made at finite distances via the Ori-Sela pre-factor. 

\begin{figure}
\includegraphics[width=7.5cm]{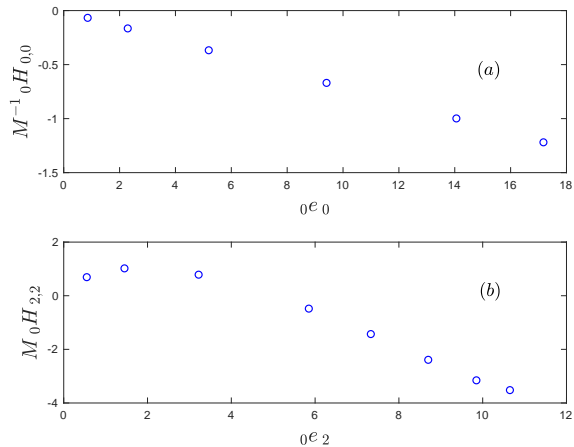}
\caption{ Top panel (a): $_0H_{0,0}$ v. $_0e_0$. We find $M^{-1}\,{_0H_{0,0}}=(0.07075\pm 0.00045)\,_0e_0 -(0.0034\pm 0.0045)$. 
Bottom panel (b): $_0H_{2,2}$ v. $_0e_2$. We find $M\,{_0H_{2,2}}=(-0.647\pm 0.058)\,_0e_2+(3.29\pm 0.50)$ (from the linear part of the figure).}
\label{000_022}
\end{figure}

\begin{figure}
\includegraphics[width=7.5cm]{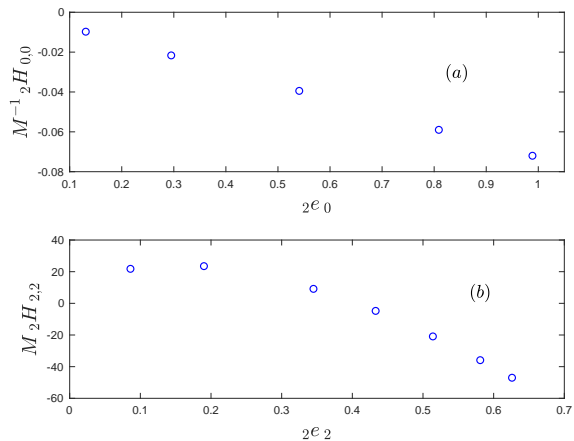}
\caption{Top panel (a): $_2H_{0,0}$ v. $_2e_0$. We find $M^{-1}\,{_2H_{0,0}}=(-0.07259\pm 0.00017)\,_2e_0 -(2.34\pm 1.08)\times 10^{-4}$. 
Bottom panel (b): $_2H_{2,2}$ v. $_2e_2$. We find $M\,{_2H_{2,2}}=(-218.\pm 31.)\,_2e_2+(90.\pm 17.)$ (from the linear part of the figure).
}
\label{200_222}
\end{figure}

\begin{figure}
\includegraphics[width=7.5cm]{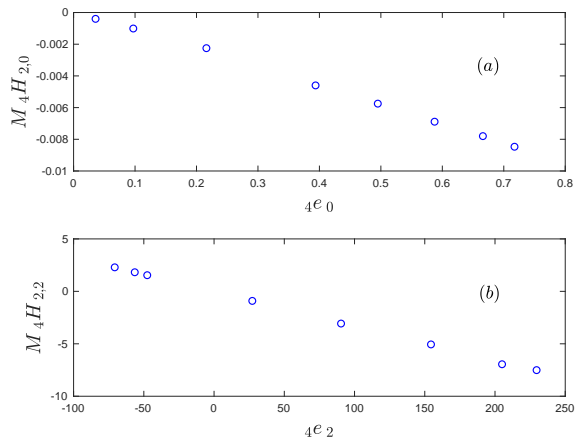}
\caption{Top panel (a): $_4H_{2,0}$ v. $_4e_0$. We find $M\,{_4H_{2,0}}=(-0.01195\pm 0.00034)\,_4e_0 -(1.46\pm 1.60)\times 10^{-4}$. 
Bottom panel (b): $_4H_{2,2}$ v. $_4e_2$. We find $M\,{_4H_{2,2}}=(-0.03297\pm 0.00064)\,_4e_2 -(0.038\pm 0.085)$.}
\label{420_422}
\end{figure}

\section{Concluding Remarks}

We show that Aretakis charges on $\mathscr{H}^+$ for extreme Kerr BH with axisymmetric scalar field perturbations  are associated with generalized Ori-Sela prefactors that are measured at finite distances.  For all cases studied we find a linear relationship of the two quantities when the initial data sets are in the near field. This relationship suggests that one could at least in principle measure the generalized Ori-Sela prefactor at a finite distance, and infer on the associated Aretakis charge on $\mathscr{H}^+$. If robust, this procedure would violate the no-hair theorems \cite{Bekenstein:1972}  in this sense. 

The cases that lead to deviation from linearity for initial data sets that are farther away from the EH warrant further investigation, possibly using stronger computational resources than those currently available to us. Our proposal regarding the role played by subdominant modes can be investigated with the case $\ell^{\prime}=0, \ell=4$ which is a subdominant mode for $_0H_{2,2}$. 

It is currently not known whether the linearity found for the relationship of the Aretakis charges and the generalized Ori-Sela prefactors are specific for axisymmetric modes of a linearized scalar field, or whether they extend also to non-axisymmetric modes. 

The question of extending our work to gravitational perturbations of extreme Kerr spacetimes is of much interest, and awaits further study, as of the question of the fully nonlinear theory, where analogous results may be of transient nature. 

\section*{Appendix A}

Specific examples for the value of $n$ are given in Tables \ref{T2} and \ref{T3}. 
\begin{table}[h]
\begin{tabular}{|c|c|c|c|}
\hline
$\ell'=0$ & $\ell=0$ & $\ell=2$ & $\ell=4$ \\
$\ell'=2$ & &  & \\
 \hline \hline
 $\psi$ & 1 & 3 & 5 \\
\hline
 $\,\partial_u\psi$ & {\bf 0} & 2 & 4 \\
 \hline
  $\,\partial_u^2\psi$ & -1 & 1 & 3 \\
  \hline
  $\,\partial_u^3\psi$ & -2 & {\bf 0} & 2 \\
  \hline
   $\,\partial_u^4\psi$ & -3 & -1 & 1 \\ 
   \hline
\end{tabular}
\caption{The value of the power-law indices $n$ for the field $\psi$ and its transverse derivatives $\,\partial_u^m\psi\sim v^n$ for $m=0,1,2,3,4$ ($m=0$ corresponds to the field $\psi$ itself.) Here, $\ell'=0$ or $\ell'=2$, and there are horizon data. The boldfaced values correspond to Aretakis constants: $_0H_{0,0}$, $_0H_{2,2}$, $_2H_{0,0}$, and $_2H_{2,2}$.}
\label{T2}
\end{table}

\begin{table}[h]
\begin{tabular}{|c|c|c|c|}
\hline
$\ell'=4$ & $\ell=0$ & $\ell=2$ & $\ell=4$ \\
 \hline \hline
 $\psi$ & 3 & 3 & 5 \\
\hline
 $\,\partial_u\psi$ & 2 & 2 & 4 \\
 \hline
  $\,\partial_u^2\psi$ & 1 & 1 & 3 \\
  \hline
  $\,\partial_u^3\psi$ & {\bf 0} & {\bf 0} & 2 \\
  \hline
   $\,\partial_u^4\psi$ & -1 & -1 & 1 \\ 
   \hline
\end{tabular}
\caption{The value of the power-law indices $n$ for the field $\psi$ and its transverse derivatives $\,\partial_u^m\psi\sim v^n$ for $m=0,1,2,3,4$ ($m=0$ corresponds to the field $\psi$ itself.) Here, $\ell'=4$, and there are horizon data. The boldfaced values correspond to Aretakis constants: $_4H_{2,0}$ and $_4H_{2,2}$.}
\label{T3}
\end{table}

\newpage

\section*{Appendix B}

We calculate the slope and intercept of the least squares regression lines $_{\ell^{\prime}}H_{k,\ell}=\beta\,_{\ell^{\prime}}e_{\ell'}+\alpha+\epsilon_i$ with $t$-confidence intervals for $95\%$ confidence level. Here, $\epsilon_i$ are the regression residuals of the $n$ data points. We first find the standard error for the slope, $$s_{\hat\beta}=\sqrt{\frac{\sum_{i=1}^{n}\epsilon_i^2}{(n-2)\sum_{i=1}^{n}(e_i-{\bar e_i})^2}}\, ,$$ 
where $e_i$ are short notation for the Ori-Sela prefactors for the $n$ data points.  We then find the standard error for the intercept, $s_{\hat\alpha}=s_{\hat\beta}\,\sqrt{\frac{1}{n}\,\sum_i e_i^2}$ Then, the margins of error for the slope and the intercept are respectively given by 
$$\delta_{\beta}=s_{\hat\beta}\, t^{*}_{n-2}$$
and
$$\delta_{\alpha}=s_{\hat\alpha}\, t^{*}_{n-2}\, ,$$
were $t^*$ is the critical value for $n-2$ degrees of freedom. 

\begin{table}[h]
\begin{tabular}{|c|c|c|c|}
\hline
$\ell',\ell,k$ & $\beta$ & $\alpha$ & dof \\
 \hline \hline
 $0,0,0$ & $-0.07075\pm 0.00045$ & $-0.0034\pm 0.0045$ & 4 \\
\hline
 $0,2,2$ & $-0.647\pm 0.058$ & $3.29\pm 0.50$ & 3 \\
 \hline
  $2,0,0$ & $-0.07259\pm 0.00017$ & $-0.00023\pm 0.00011$ & 3 \\
  \hline
  $2,2,2$ & $-218.\pm 31.$ & $90.\pm 17.$& 2 \\
  \hline
   $4,0,2$ & $-0.01195\pm 0.00034$ & $0.00015\pm 0.00016$ & 6 \\ 
   \hline
   $4,2,2$ & $-0.03297\pm 0.00064$ & $-0.038\pm 0.085$ & 6 \\
   \hline
\end{tabular}
\caption{The $95\%$ $t-$confidence intervals for the coefficient $\beta$ (slope) and $\alpha$ (intercept) for the regression expression $_{\ell^{\prime}}H_{k,\ell}=\beta\,_{\ell^{\prime}}e_{\ell'}+\alpha$. Here, dof is the number of $t-$statistics degrees of freedom.}
\label{T5}
\end{table}

In Table \ref{T5} we show the slope and intercept coefficients for the six cases we study. 

\section*{Acknowledgements} 
The authors acknowledge support from NSF Grants No.~PHY-2106755 and DMS-1912716 (G.K). Simulations were performed on the UMass-URI UNITY supercomputer and MIT’s SuperCloud supported by the Massachusetts Green High Performance Computing Center (MGHPCC).


\begin{thebibliography}{99}

\bibitem{Couch_Torrence:1984} W.E.~Couch and R.J.~Torrence, Gen. Rel. Grav. {\bf 16}, 789-792 (1984).
\bibitem{Bizon:2013} P.~Bizo\'{n}  and H.~Friedrich, Class.~Quant.~Grav.~{\bf 30}, 065001 (2013) [arXiv:1212.0729 [gr-qc]]. 
\bibitem{Lucietti:2013} J.~Lucietti, K.~Murata, H.S.~Reall and N.~Tanahashi, JHEP {\bf 1303}, 035 (2013) [arXiv:1212.2557 [gr-qc]].
\bibitem{Godazgar:2017} H.~Godazgar, M.~Godazgar and C.N.~Pope, Phys.~Rev.~D {\bf 96}, 084055 (2017)  [arXiv:1707.09804 [Angelopoulos:2020hep-th]].
\bibitem{Bhattacharjee:2018} S.~Bhattacharjee, B.~Chakrabarty, D.D.K.~Chow, P.~Paul, and A.~Virmani, arXiv:1805.10655  (2018). 
\bibitem{Bekenstein:1972} J.D.~Bekenstein, Phys.~Rev.~Lett.~{\bf 28}, 452 (1972); Phys.~Rev.~D {\bf 5}, 1239 (1972); {\it ibid.}~2403 (1972). 
\bibitem{Angelopoulos:2018} Y.~Angelopoulos, S.~Aretakis, and D.~Gajic, Phys.~Rev.~Lett.~{\bf 121}, 131102 (2018)
\bibitem{Burko:2019} L.M.~Burko, G.~Khanna, and S.~Sabharwal, Phys.~Rev.~Research {\bf 1}, 033106 (2019) [arXiv:1906.03116 [gr-qc]]. 
\bibitem{Burko:2021}  L.M.~Burko, G.~Khanna, and S.~Sabharwal, Phys.~Rev.~D {\bf 103}, 021502 (2021) [arXiv:2005.07294 [gr-qc]]. 
\bibitem{Burko:2014} L.M.~Burko and G.~Khanna, Phys.~Rev.~D {\bf 89}, 044037 (2014) [arXiv:1312.5247 [gr-qc]]. 
\bibitem{Ori:2013} A.~Ori, arXiv:1305.1564 [gr-qc] (2013).
\bibitem{Sela:2016} O.~Sela, Phys.~Rev.~D {\bf 93}, 024054 (2016) [arXiv:1510.06169 [gr-qc]]. 
\bibitem{code} Scott E. Field, Sigal Gottlieb, Zachary J. Grant, Leah F. Isherwood, Gaurav Khanna, Commun. Appl. Math. Comput. 5, 97–115 (2023). https://doi.org/10.1007/s42967-021-00129-2.
\bibitem{Angelopoulos:2020} Y.~Angelopoulos, S.~Aretakis, and D.~Gajic, Adv.~Math.~{\bf 375}, 107363 (2020) [arXiv:1807.03802].






\end{thebibliography}
\end{document}